\documentclass[letterpaper, 10 pt, conference]{ieeeconf}  % Comment this line out if you need a4paper

\IEEEoverridecommandlockouts                              % This command is only needed if 
\usepackage{amsmath} % assumes amsmath package installed
\usepackage{amssymb}  % assumes amsmath package installed
\usepackage{caption}
\usepackage{subcaption}
\usepackage{graphicx}
\usepackage{algorithm}
\usepackage{algpseudocode}
\usepackage[utf8]{inputenc}
\usepackage{hyperref}
\usepackage{cite} % For citation commands
\providecommand{\keyword}[1]
{
  \small	
  \textbf{\textit{Keywords---}} #1
}

\title{\LARGE \bf
Decoupled Thrust-Axis Attitude Control Using Quaternions for Chandrayaan-3 Lunar Landing Mission}
\author{Aditya Rallapalli$^{1}$, Suraj Kumar$^{1}$, Rijesh M P $^{1}$ Ashok Kumar Kakula$^{1}$, Bharat Kumar GVP$^{1}$
\thanks{$^{1}$ Controls and Digital Area, U R Rao Satellite Centre, Indian Space Research Organization (ISRO), Bangalore, India (e-mail: \{adityar,surajk,kashok,bharat\}@ursc.gov.in)%
}
}

\begin{document}

\maketitle
\thispagestyle{empty}
\pagestyle{empty}

%%%%%%%%%%%%%%%%%%%%%%%%%%%%%%%%%%%%%%%%%%%%%%%%%%%%%%%%%%%%%%%%%%%%%%%%%%%%%%%%
\begin{abstract}
India’s Chandrayaan-3 mission achieved a historic milestone with its successful soft landing near the lunar south pole, highlighting the critical role of the navigation, guidance, and control (NGC) system. Navigation provided vehicle state estimates relative to the Moon’s center, while a polynomial-based guidance scheme computed the required acceleration profile to meet terminal landing conditions. This acceleration demand was translated into total thrust magnitude and attitude commands generation.
Attitude command generation involved aligning the thrust axis with the required acceleration vector and constraining rotation about the thrust axis, typically governed by mission-specific requirements. Although quaternion-based control laws are preferred for their singularity-free representation, they inherently couple all three rotational axes. This coupling can lead to undesirable interactions between guidance and control, especially during large rotations about the thrust axis, due to the quaternion’s shortest-path property.
This paper proposes a novel quaternion-based decoupling method that enables independent thrust-axis control, mitigating guidance-control interaction and ensuring proper attitude commands generation for lander attitude control.    
\end{abstract}
\keyword{\small Lunar lander, Attitude control system, Thrust axis control}

%%%%%%%%%%%%%%%%%%%%%%%%%%%%%%%%%%%%%%%%%%%%%%%%%%%%%%%%%%%%%%%%%%%%%%%%%%%%%%%%
\section{INTRODUCTION}
India’s Chandrayaan-3 mission marked a significant technological milestone by achieving a successful soft landing near the lunar south pole—an area of strategic scientific interest and operational challenge. This accomplishment demonstrated India's advancement in autonomous planetary landing systems and placed it among a small group of nations with demonstrated capability for precision lunar descent and touchdown.

A key enabler of this success was the onboard autonomous navigation, guidance, and control (NGC) system. The autonomous NGC system was activated at perilune, approximately 30 km above the lunar surface. The powered descent phase was facilitated by four 800 N throttleable liquid engines, which performed the necessary deceleration, complemented by eight 58 N attitude control thrusters used to orient the lander along the required direction. The entire powered descent phase was divided into multiple subphases to ensure robust and reliable landing performance under varying conditions\cite{rallapalli2024landing}\cite{rallapalli2025onboard}\cite{aditya2024stability}.

During the powered descent phase, navigation provided accurate estimates of the vehicle’s position, velocity, and attitude relative to the Moon’s center. Based on these states and the available time-to-go, the onboard polynomial-based guidance algorithm \cite{rijeshdesign}\cite{chakrabarti2024convex} generated the required acceleration commands to ensure that terminal conditions—such as landing velocity and altitude—were met. 

To track these acceleration commands, the control system resolved the required acceleration to the total thrust magnitude and the desired/reference attitude of the vehicle. The attitude reference generation process typically involves two primary constraints: (1) aligning the thrust axis with the required acceleration vector, and (2) constraining rotation about the thrust axis to satisfy mission-specific sensor orientation or communication requirements. While the on-board guidance law does not impose a strict constraint on rotation about the thrust axis, its alignment is often indirectly dictated by mission constraints such as optical sensor visibility or antenna pointing. It is therefore preferable to use a decoupled controller to address these two requirements independently. 

Quaternion-based attitude control laws are widely adopted in aerospace applications because of their ability to represent rotations without singularities. However, quaternion feedback inherently couples all three rotational degrees of freedom. This coupling becomes problematic when large rotations about the thrust axis are required to align certain sensors in a specific area of interest, as the quaternion controller naturally follows the shortest rotation path. This can result in unintended interactions between the guidance commands and the control response, potentially degrading tracking performance, and can lead to instability if control bandwidth is not sufficiently higher than guidance bandwidth.

Quaternion-based attitude control laws are extensively studied in the literature\cite{sidi1997spacecraft}. In \cite{johnson2006parameterized}, a reduced-quaternion-based linear quadratic regulator (LQR) and sliding mode controller were presented for thrust vector control. In \cite{kwon2016virtual}, a sequential force-torque control approach was proposed to address guidance and control interaction, enabling touchdown with a desired descent path angle using dual quaternion kinematics. However, the proposed design assumes no rotation is required about the thrust axis. In practice, significant rotation about the thrust axis may be necessary to orient optical sensors—such as hazard detection cameras—toward sunlit regions \cite{amitabh2023terrain}.

To address this limitation, this paper presents an analytical method to decouple thrust-axis rotation from lateral-axis control, thereby enabling each to independently meet mission-specific requirements and guidance demands. The remainder of the paper is organized as follows: Section II describes the translational and rotational dynamics that govern the motion of the lander near the lunar surface. Section III details the computation of the reference attitude based on the acceleration demand generated by the polynomial guidance algorithm. Section IV introduces the proposed decoupled attitude control methodology, designed to independently satisfy both mission-specific and guidance-related requirements. Section V presents a comparative analysis between the proposed decoupled control scheme and the classical quaternion-based attitude controller. Finally, Section VI summarizes the key findings and contributions of the work. 

\section{Lunar Lander Dynamics}  
\label{sec:prw}
The motion of a lunar lander is governed by six-degree-of-freedom (6-DoF) equations, which capture the coupled translational and rotational dynamics necessary for accurate modeling during descent and landing phases. Translational dynamics can be expressed in moon centered moon fixed reference frame using spherical co-ordinate reference is given by \eqref{traj_opt_dyn}. However, at close proximity with non-rotating flat moon approximation and significant
velocity reduction, the Coriolis and moon’s rotational velocity term in dynamics can be neglected as their contribution to total acceleration is negligible \eqref{lander_dynamics4}

\begin{figure}[t]
  \centering
  \includegraphics[width=0.5\textwidth]{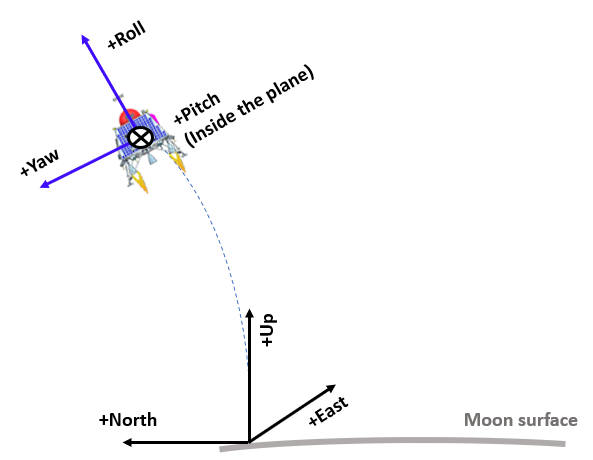}
  \caption{Frame of reference used for navigation, guidance and control}
  \label{fig:ref_frame}
\end{figure}
\begin{align}
\label{traj_opt_dyn}
    \dot{r} &= w \nonumber \\
    \dot{\theta} &= \frac{u}{rcos\phi} \nonumber \\
    \dot{\phi} &= \frac{v}{r} \nonumber \\
    \dot{w}  &= \frac{Tsin\beta}{m} - \frac{\mu}{r^2}  + \frac{(u^2 + v^2)}{r} \\ 
    &+ (-2u\omega_m cos\phi + r\omega_m^2 cos^2\phi) \nonumber \\
    \dot{u} &= \frac{Tcos\alpha cos\beta}{m} + \frac{(-uw + uv tan\phi)}{r} \\
    &+ (-2w\omega_m cos\phi + 2v\omega_m sin\phi) \nonumber \\
    \dot{v} &= \frac{(Tsin\alpha cos\beta)}{m} \frac{(-vw - u^2 tan\phi)}{r} \\
    &+ (-2u\omega_m sin\phi - r\omega_m^2 sin\phi cos\phi) \nonumber \\
    \dot{m} &= -\frac{T}{I_{sp}g_0}
\end{align}
Here, $r$ represents the radial distance from the moon centre, $\theta, \phi$ represents the lander latitude and longitude respectively; $\omega_m$ represents moon rotational velocity; $u,v,w$ represents the tangential, across and radial velocity components respectively; $m$ represents the mass of the lander; $\mu$ represents the moon's gravitational constant; $I_{sp}$ represents specific impulse of the engine and $g_0 = 9.81$. The thrust vector is parameterized in the body frame by its magnitude $T$, declination angle, $\beta$, and right ascension angle $\alpha$. 

\begin{eqnarray}
&\dot{\boldsymbol{{r}}} &=\quad \boldsymbol{{v}} \nonumber \\
&\boldsymbol{\dot{{v}}} &=\quad \boldsymbol{{g}} + \boldsymbol{{a}} \nonumber \\
&||{a}|| &=\quad \frac{T}{m}   \nonumber \\
&\dot{m} &=\quad -\frac{T}{I_{sp} g_{0}}  \label{lander_dynamics4}
\end{eqnarray}

where ${\boldsymbol{r} \in \mathbb{R}^3}, {\boldsymbol{v} \in \mathbb{R}^3}$ and ${\boldsymbol{a} \in \mathbb{R}^3}$ are the position, velocity and acceleration vectors expressed in the local reference frame, i.e. North, Up and East (NUE); $\boldsymbol{{g}}$ is the moon gravity vector; $I_{sp}$ is the specific impulse of engine. 

Rotational dynamics is defined using euler moment equation and quaternion kinematics \eqref{lander_rotdynamics4},
\begin{eqnarray}
&{\boldsymbol{I\dot{\omega}}} + \boldsymbol{\omega} \times \boldsymbol{I\omega} &=\quad \boldsymbol{{\tau_{RCS}} + \boldsymbol{\tau_{Engine}}} \nonumber \\
& \dot{\mathbf{q}} &= \frac{1}{2} \boldsymbol{\Omega(\omega)}\mathbf{q} \nonumber \\
& \boldsymbol{\Omega(\omega)} &= 
    \begin{bmatrix} 
    0 & \omega_z & -\omega_y & \omega_x \\
    -\omega_z & 0 & \omega_x & \omega_y \\
    \omega_y & -\omega_x & 0 & \omega_z \\
    -\omega_x & \omega_y & -\omega_z & 0 \\   
\end{bmatrix}
\label{lander_rotdynamics4}
\end{eqnarray}

where $ (q \in \mathbb{R}^4 \mid ||q||_2 = 1)$ is the attitude of lander with respect to local frame, ${\omega(\omega_x,\omega_y,\omega_z ) \in \mathbb{R}^3}$ and ${\tau_{RCS} , \tau_{Engine}(\tau_x,\tau_z,\tau_z) \in \mathbb{R}^3}$ are angular velocity and moment due to RCS and engines expressed in body frame, i.e. yaw, roll and pitch. Fig.\ref{fig:ref_frame} represent the frame of reference used for equation of motion.

\section{Reference Attitude Command}
The primary role of the guidance system during the powered descent phase is to compute the required translational acceleration that will steer the lander from its current state to the desired terminal conditions. These boundary conditions are specified in terms of position, velocity, and sometimes acceleration constraints to maintain phase continuity. In order to meet these boundary condition we represent the net guidance acceleration, ${\bar{a}}(t)$, using a third-order polynomial \eqref{eq3}
\begin{equation}
\label{eq3}
    \boldsymbol{\bar{a}}(t) = \boldsymbol{C}_{0} + \boldsymbol{C}_{1}t + \boldsymbol{C}_{2}t^2 + \boldsymbol{C}_{3}t^3
\end{equation}
Co-efficients of the cubic polynomial i.e. $C_0,C_1,C_2$ and $C_3$ are arrived by solving two point boundary value problem as
\begin{equation}
\label{gui_lin_eq}
    \begin{bmatrix} 
    1 & 0 & 0 & 0 \\
    1 & t_{\text{go}} & t_{\text{go}}^2 & t_{\text{go}}^3 \\
    t_{\text{go}} & \frac{t_{\text{go}}^2}{2} & \frac{t_{\text{go}}^3}{3} & \frac{t_{\text{go}}^4}{4} \\
    \frac{t_{\text{go}}^2}{2} & \frac{t_{\text{go}}^3}{6} & \frac{t_{\text{go}}^4}{12} & \frac{t_{\text{go}}^5}{20} \\
\end{bmatrix}\begin{bmatrix} C_{0}^i \\ C_{1}^i \\ C_{2}^i \\ C_{3}^i \end{bmatrix} = \begin{bmatrix}
\boldsymbol{a}_0^i \\
\boldsymbol{a}_f^i \\
\boldsymbol{v}_f^i - \boldsymbol{v_0}^i \\
\boldsymbol{r}_f^i - \boldsymbol{r_0}^i - \boldsymbol{v_0}^i \cdot t_{\text{go}} \\
\end{bmatrix} \\
\end{equation}
$\mathbf{r}_0$, $\mathbf{v}_0$, $\mathbf{a}_0$ represent the initial position, velocity, and acceleration, respectively, and $\mathbf{r}_f$, $\mathbf{v}_f$, $\mathbf{a}_f$ represent the terminal position, velocity, and acceleration. Index $i$ represents the indices of the $x$, $y$, $z$ components of the vector. Based on current $t_{go}$ which is decremented linearly once guidance is initiated, acceleration required in current step time is computed as
\begin{equation}
    \boldsymbol{\bar{a}} = \boldsymbol{C}_{0} + \boldsymbol{C}_{1}g_{step} + \boldsymbol{C}_{2}g_{step}^2 + \boldsymbol{C}_{3}g_{step}^3
\end{equation} 
where $g_{step}$ is the guidance cycle time. \\

For phases like hovering where primary aim is to maintain current position, guidance acceleration is derived based on position and velocity guidance defined by

\begin{equation}
    \boldsymbol{\bar{a}} = \boldsymbol{g} + K_r \boldsymbol{r_e}+K_v\boldsymbol{v_e}
\end{equation} 
where $\boldsymbol{r_e} = \boldsymbol{r_{ref}} - \boldsymbol{r}$ is the position error,$\boldsymbol{v_e}= \boldsymbol{v_{ref}} - \boldsymbol{v}$ is velocity error, $K_r$ and $K_v$ are positive-definite proportional and derivative guidance gain   \\

The acceleration command generated by the guidance system must be realized through appropriate thrust vectoring and control of the thrust magnitude. The required thrust magnitude is computed based on the current mass of the lander and the norm of the commanded acceleration vector, as given by
\begin{equation}
    {T} = m ||a||_2
\end{equation}

To ensure that the generated thrust produces the desired acceleration, the reference attitude must be computed such that the thrust vector aligns with the direction of the commanded acceleration via the shortest possible rotation. This ensures smooth attitude transitions and avoids undesirable trajectory deviations due to guidance and control interactions. Let the thrust direction in the body frame be defined as
\begin{equation}
    \boldsymbol{\hat{t}} = [0,1,0]^T
\end{equation}

This unit vector represents the direction of the engine thrust in the body-fixed frame. In this formulation, it is assumed that the engines generate thrust along the positive roll axis of the lander.

Define normalized required acceleration unit vector as
\begin{equation}
    \boldsymbol{\hat{a}} = \frac{\boldsymbol{a}}{||a||_2}
\end{equation}

Then, reference attitude which continuously maps thrust direction to required acceleration direction in shortest possible path is given by,
\begin{equation}
\begin{aligned}
       \boldsymbol{\hat{n}} &= \boldsymbol{\hat{t}} \times \boldsymbol{\hat{a}} \nonumber &&     
    \theta &= cos^{-1}(\boldsymbol{\hat{t}}.\boldsymbol{\hat{n}}) \nonumber \\
    \mathbf{q}_{guid} &= \begin{bmatrix}
        cos(\frac{\theta}{2}) \\
        \boldsymbol{\hat{n}}sin(\frac{\theta}{2})       
    \end{bmatrix}
\end{aligned}
\end{equation}

\noindent where:
\begin{itemize}
    \item $\hat{n} \in \mathbb{R}^3$ is axis of rotation
    \item $\theta \in \mathbb{R}$ is angle of rotation
  \item \( \mathbf{q} = [q_0,\ \mathbf{q}_v] \in \mathbb{R}^4 \) is a quaternion represented as:
    \begin{itemize}
        \item \( q_0 \in \mathbb{R} \): scalar part
        \item \( \mathbf{q}_v \in \mathbb{R}^3 \): vector part
        \item subscript guid represent guidance generated
    \end{itemize}
\end{itemize}
Above mentioned control law ensures shortest path is travelled to meet the guidance orientation requirement.
Let mission specific constraint to point optical sensor to a certain area of interest requires a $\phi_{cmd}$ rotation about thrust axis. The attitude reference for mission specific constraint is formulated as
\begin{equation}
\begin{aligned}
A[\phi_{cmd}] &= \begin{bmatrix}
    cos(\phi_{cmd}) & 0 &-sin(\phi_{cmd}) \\
      0  & 1 & 0 \\
      sin(\phi_{cmd}) & 0 & cos(\phi_{cmd})
\end{bmatrix} \\
\mathbf{q}_{\phi cmd} &\gets A[\phi_{cmd}] 
\end{aligned}
\end{equation}
where $A[\theta]$ is a direction cosine matrix $A[\theta] \in SO(3)$ and $q_{\phi cmd}$ is quaternion obtained from direction cosine matrix using standard closed form solution \cite{sidi1997spacecraft}.
Attitude reference generated via guidance and mission specific requirements is obtained as
\begin{equation}
    \mathbf{q}_{ref} = \mathbf{q}_{guid} \otimes \mathbf{q}_{\phi cmd}
\end{equation}

\section{Decoupled Thrust Axis Control}
The development and analysis of quaternion-based attitude control laws have been thoroughly addressed in existing research. A linear proportional plus derivative attitude control law using quaternion is formulated.
The attitude tracking error is represented by the error quaternion \( \mathbf{q}_e \), computed as:
\[
\mathbf{q}_e =
\begin{bmatrix}
q_{e0} \\
\mathbf{q}_{ev}
\end{bmatrix}
=
\mathbf{q}_{ref}^{-1} \otimes \mathbf{q}
=
\begin{bmatrix}
q_{ref0} q_0 + \mathbf{q}_{refv}^\top \mathbf{q}_v \\
q_{ref0} \mathbf{q}_v - q_0 \mathbf{q}_{refv} + \mathbf{q}_{refv} \times \mathbf{q}_v
\end{bmatrix}
\]
The control torque \( \boldsymbol{\tau_{cmd}} \in \mathbb{R}^3 \) is computed using a quaternion-based PD control law:

\[
\boldsymbol{\tau_{cmd}} = -K_p \, \mathbf{q}_{ev} - K_d \, \boldsymbol{\omega}_e + \boldsymbol{\omega} \times \boldsymbol{I}\boldsymbol{\omega}
\]

\noindent where:
\begin{itemize}
  \item \( \boldsymbol{\tau_{cmd}} \in \mathbb{R}^3 \): commanded control torque
  \item \( \boldsymbol{\omega}_e = \boldsymbol{\omega} - \boldsymbol{\omega_{ref}} \in \mathbb{R}^3 \): angular velocity error
  \item \( \boldsymbol{\omega} \): current body angular velocity
  \item \( \boldsymbol{\omega}_{ref} \): desired angular velocity 
  \item \( K_p, K_d \in \mathbb{R}^{3 \times 3} \): positive-definite proportional and derivative gain matrices
\end{itemize}

\begin{figure}[t]
  \centering
  \includegraphics[width=0.5\textwidth]{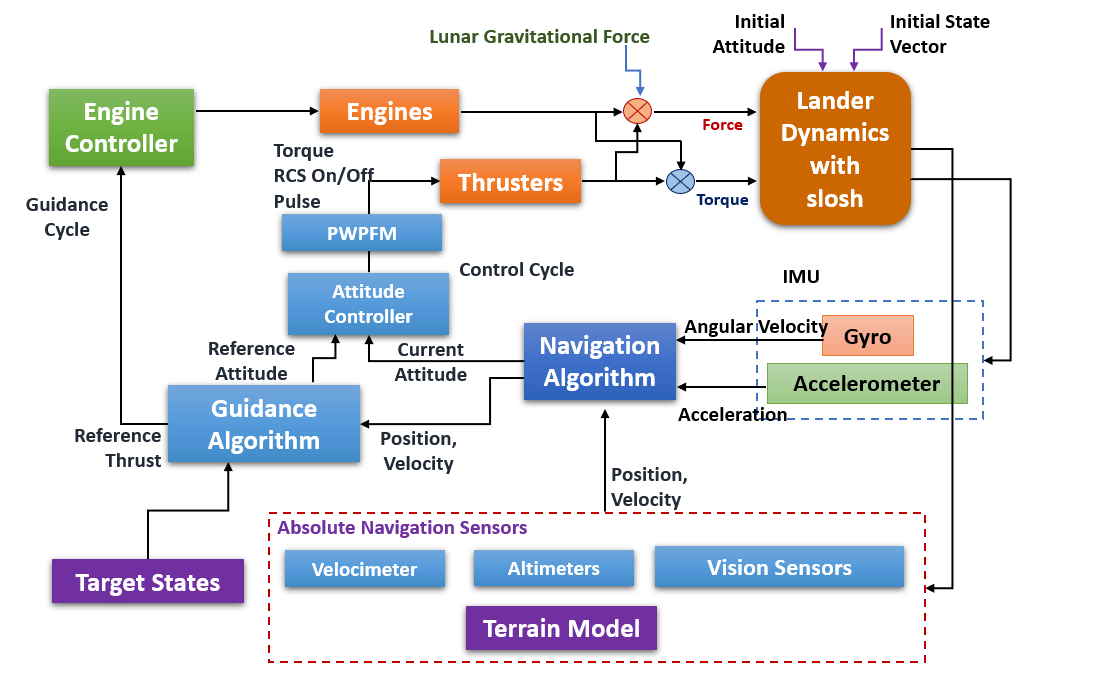}
  \caption{Navigation, guidance and control architecture for lunar landing mission}
  \label{fig:ngc_block}
\end{figure}

Fig.\ref{fig:ngc_block} represent the overall working of navigation, guidance and control system described till now. 
Fig. \ref{fig:issue} illustrates a representative scenario with two distinct cases analyzed. Consider an initial condition where the thrust vector is aligned at an angle of $15^0$ from the local vertical, oriented toward the north, and the pitch axis is aligned along the east direction.
\begin{itemize}
    \item In Case 1, when the commanded rotation about the thrust axis is $\phi_{cmd} = 0^0$ , the guidance system requires reorientation of the thrust vector to align with the vertical direction. The attitude controller interprets this as a pure pitch rotation and selects the pitch axis as the axis of rotation. Consequently, the controller generates a smooth, shortest-path rotation that aligns the thrust vector with the vertical direction, as desired by the guidance, resulting in no guidance-control interaction.
    \item In Case 2, when $\phi_{cmd} = 180^0$, the desired attitude introduces a large rotation about the thrust axis. The coupled quaternion-based attitude controller selects an eigen axis close to the roll axis for the shortest-path rotation. As a result, the thrust vector initially moves toward the east, deviating from the vertical path intended by the guidance system. This deviation introduces an undesirable motion component not aligned with the guidance direction, potentially leading to degraded system performance due to guidance-control interaction.
\end{itemize}
To mitigate the above mentioned issue, we propose a decoupled thrust-axis attitude controller. In this approach, attitude error for lateral axis and thrust axis is computed independently based on guidance and mission constraints. Achieved rotation about thrust axis $\phi_b$ is computed from current body quaternion based on euler angles with 1-3-2 rotation sequence. Guidance generated reference is then augmented with current thrust axis rotation to generate decouple lateral axis control. Thrust axis control is then performed based on required $\phi_{cmd}$ and current $\phi_b$.
In a 1-3-2 Euler rotation sequence, the composite rotation matrix is derived by sequentially rotating about the \( x \)-axis by angle \( \phi \), then the \( z \)-axis by angle \( \psi \), and finally about the new \( y \)-axis by angle \( \theta \). Using shorthand notations:
\begin{figure*}[t]
  \centering
  \includegraphics[width=0.8\textwidth]{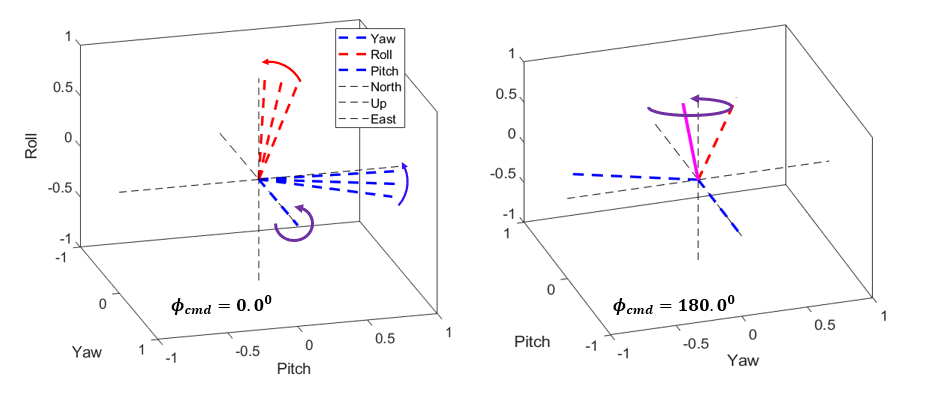}
  \caption{Guidance and control interaction (i) shows the guidance demand being met without mission constraints (ii) shows deviation in guidance demand as quaternion tries to achieve shortest path to achieve both mission and guidance demand simultaneously}
  \label{fig:issue}
\end{figure*}
the direction cosine matrix \( R_{1\text{-}3\text{-}2} \) can be expressed as:

\begin{equation}
 R =
\begin{bmatrix}
c_\theta c_\psi & c_\theta s_\psi c_\phi + s_\theta s_\phi & c_\theta s_\psi s_\phi - s_\theta c_\phi \\
-s_\psi & c_\psi c_\phi & c_\psi s_\phi \\
s_\theta c_\psi & s_\theta s_\psi c_\phi - c_\theta s_\phi & s_\theta s_\psi s_\phi + c_\theta c_\phi
\end{bmatrix}
\end{equation}
\begin{itemize}
    \item {$c_{\theta}$} is $cos(\theta)$
    \item {$s_{\theta}$} is $sin(\theta)$
    \end{itemize}   
Based on direction cosine matrix generated using current quaternion $\mathbf{q}$, $\phi_b$ can be derived analytically as

\begin{equation}
\phi_b = \tan^{-1}(\frac{2(q_1q_3+q_2q_0)}{q_1^2-q_2^2+q_3^2+q_4^2})
\end{equation}

\noindent where the quaternion is defined as:
\[
\mathbf{q} = 
\begin{bmatrix}
q_0 \\
\mathbf{q}_v 
\end{bmatrix}
\]
Attitude reference to decouple lateral axis from thrust axis is given by
\begin{equation}
\begin{aligned}
A[\phi_{b}] &= \begin{bmatrix}
    cos(\phi_{b}) & 0 &-sin(\phi_{b}) \\
      0  & 1 & 0 \\
      sin(\phi_{b}) & 0 & cos(\phi_{b})
\end{bmatrix} \\
\mathbf{q}_{\phi_b} &\gets A[\phi_{b}] \\
    \mathbf{q}_{ref_L} &= \mathbf{q}_{guid} \otimes \mathbf{q}_{\phi_b}
\end{aligned}
\end{equation}
Attitude tracking error for lateral axis is computed as
\[
\mathbf{q}_e =
\begin{bmatrix}
q_{e0} \\
\mathbf{q}_{ev}
\end{bmatrix}
=
\mathbf{q}_{refL}^{-1} \otimes \mathbf{q}
\]

The attitude tracking errors are extracted from the components of the error quaternion \( \mathbf{q}_e \). Let the components be expressed as:

\[
\mathbf{q}_e =
\begin{bmatrix}
q_{e0} \\
q_{ex} \\
q_{ey} \\
q_{ez}
\end{bmatrix}
\]

The control errors about the body axes are defined as:
\begin{equation}
\begin{aligned}
\delta\theta_r &= \phi_b - \phi_{\text{cmd}} \quad &\text{(roll-axis error)} \\
\delta\theta_y &= q_{ex} \quad &\text{(yaw-axis error)} \\
\delta\theta_p &= q_{ez} \quad &\text{(pitch-axis error)}
\end{aligned}
\end{equation}

\noindent where:
\begin{itemize}
  \item \( \delta\theta_r \) represents the roll-axis error computed from thrust-axis rotation: \( \phi_b - \phi_{\text{cmd}} \), to meet mission requirements
  \item \( \delta\theta_y \) represents the yaw-axis attitude error to meet guidance requirement,
  \item \( \delta\theta_p \) represents the pitch-axis attitude error meet guidance requirement.
\end{itemize}

\section{Results}  \label{sec:results}
To validate the effectiveness of the proposed decoupled thrust-axis attitude control strategy, extensive simulations were conducted using a high-fidelity, end-to-end autonomous trajectory simulator \cite{chaitanya2024chandrayaan}. This simulator includes detailed models of the lunar environment, vehicle dynamics, sensor noise, actuator characteristics, and onboard navigation, guidance, and control (NGC) algorithms representative of those flown on Chandrayaan-3.
Tables \ref{tab:sim_params} and Table \ref{tab:ngc_params} represent the parameters taken to simulate a typical hovering and vertical descent case where it is required to point the optical camera once we reach the required target altitude.
\begin{figure}[t]
  \centering
  \includegraphics[width=0.5\textwidth]{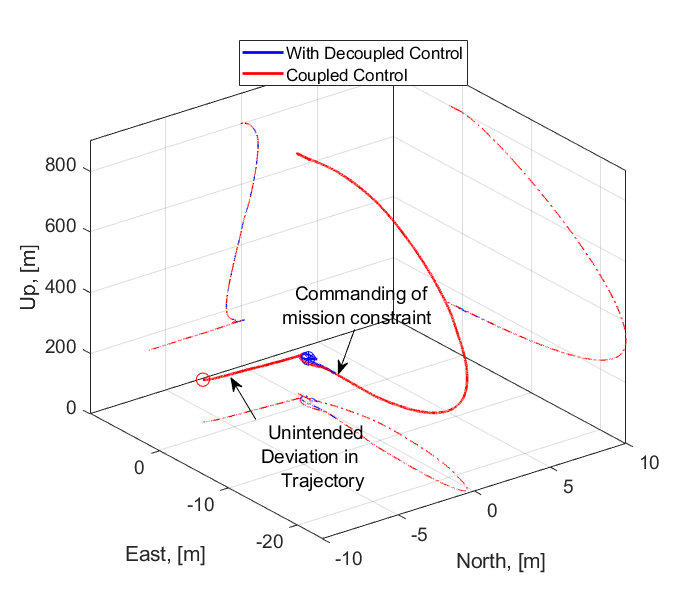}
  \caption{Comparison of descent trajectory with and without decouple attitude control}
  \label{fig:r1}
\end{figure}

\begin{figure}[t]
  \centering
  \includegraphics[width=0.5\textwidth]{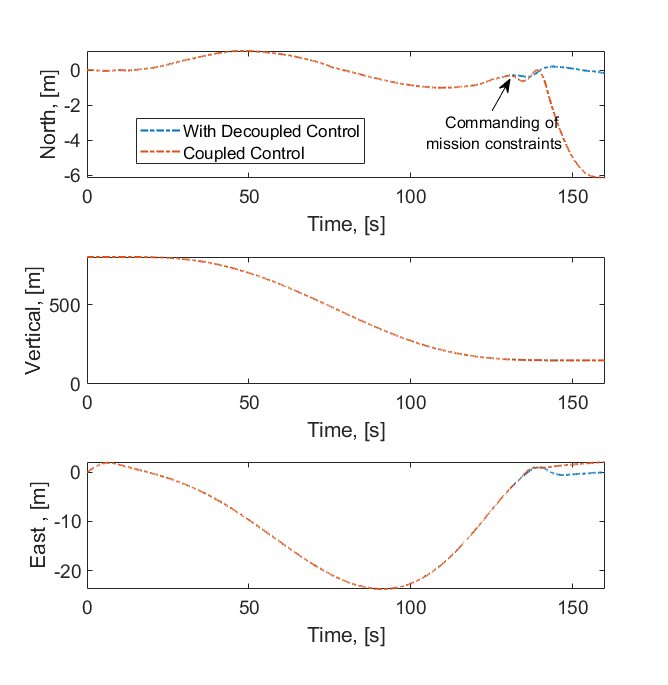}
  \caption{Time history of position for coupled and decoupled controller}
  \label{fig:r2}
\end{figure}

\begin{figure}[t]
  \centering
  \includegraphics[width=0.5\textwidth]{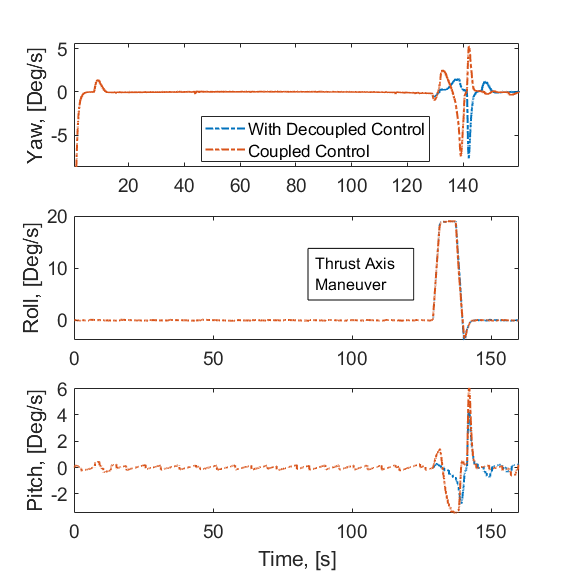}
  \caption{Angular velocity comparison for the two scenarios}
  \label{fig:r3}
\end{figure}

\begin{table}[t]
\centering
\caption{Lander Simulation Parameters}
\begin{tabular}{|l|c|}
\hline
\textbf{Parameter} & \textbf{Value} \\
\hline
Lunar gravity (\(g\)) & \(1.62 \, \text{m/s}^2\) \\
Initial mass (\(m_0\)) & \(900 \, \text{kg}\) \\
Moment of inertia (\(I_{xx}\)) & \(400 \, \text{kg} \cdot \text{m}^2\) \\
Moment of inertia (\(I_{yy}\)) & \(300 \, \text{kg} \cdot \text{m}^2\) \\
Moment of inertia (\(I_{zz}\)) & \(450 \, \text{kg} \cdot \text{m}^2\) \\
Reaction Control System & \(8 \times 58 \, \text{N}\) \\
Throttleable Engines & \(2 \times (800-360) \, \text{N}\) \\
\hline
\end{tabular}
\label{tab:sim_params}
\end{table}
\begin{table}[t]
\centering
\caption{Lander NGC Parameters}
\begin{tabular}{|l|c|}
\hline
\textbf{Parameter} & \textbf{Value} \\
\hline
Control Bandwidth (\(\omega_n\)) & \(0.5 \, \text{Hz}\) \\
Control Damping (\(\zeta_n\)) & \(0.8 \, \) \\
Guidance Bandwidth (\(\omega_n\)) & \(0.06 \, \text{Hz}\) \\
Guidance Damping (\(\omega_n\)) & \(0.8 \, \) \\
Vertical Descent Target ($\boldsymbol{r_{Ty}}$) & \(150 \,m \) \\
Initial Vertical Position ($\boldsymbol{r_{0y}}$) & \(800 \,m \) \\
Time to go ($\boldsymbol{T_{go}}$) & \(130 \,s \) \\
Mission constraints ($\boldsymbol{\phi_{cmd}}$) & \(150  \,deg \) \\
Current thrust axis rotation ($\boldsymbol{\phi_{b}}$) & \(0 \,deg \) \\
\hline
\end{tabular}
\label{tab:ngc_params}
\end{table}
\begin{figure}[t]
  \centering
  \includegraphics[width=0.5\textwidth]{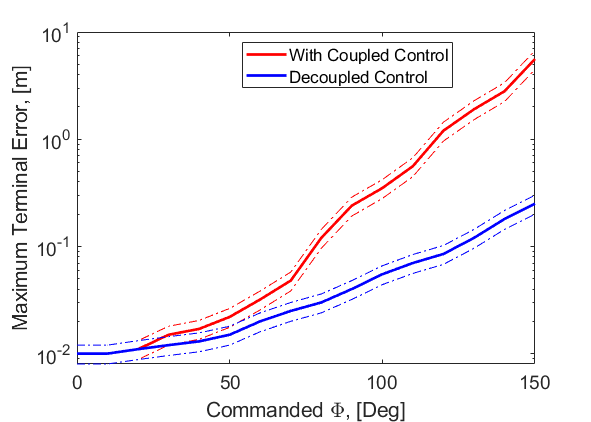}
  \caption{Monte carlo simulation based terminal error comparison of coupled and decoupled control design}
  \label{fig:r4}
\end{figure}

To evaluate the performance of the proposed decoupled controller, a representative simulation scenario was constructed. In this scenario, the lander initially performs a hover at an altitude of 800 m for 12 seconds, guided by a position–velocity (PV) based guidance scheme. This is followed by a vertical descent from 800 m to 150 m, governed by a third-order polynomial guidance law. Upon reaching 150 m altitude, the lander executes an extended hover phase lasting 30 seconds to facilitate imaging and hazard-free region identification. During the descent phase, a mission-specific thrust-axis rotation requirement of $\phi_{\text{cmd}} = 150^\circ$ is introduced to ensure that the onboard camera is correctly oriented toward the target region of interest. This is achieved via introducing an attitude maneuver about lander roll axis which can be seen in Fig.\ref{fig:r3}. 
Fig.\ref{fig:r1} illustrates the three-dimensional translational trajectories comparing the performance of the proposed decoupled controller with a conventional quaternion-based attitude controller. It is evident that upon initiating the mission-specific thrust-axis rotation requirement, the conventional controller exhibits significant deviation from the desired path. In contrast, the proposed decoupled controller maintains minimal deviation and successfully meets the terminal accuracy constraints. 
Fig.\ref{fig:r2} presents the time history of the position components, while Fig.\ref{fig:r3} shows the corresponding angular velocity profiles. Using these figures, it is evident that conventional quaternion based control law introduces large guidance and control interaction while decouple thrust axis law ensures mission and guidance requirements are met independently. A Monte Carlo analysis was performed for varying commanded values of $\phi_{\text{cmd}}$. The results indicate that for small commanded angles, the performance difference between the two controllers is negligible. However, as the thrust-axis rotation demand increases, the conventional controller demonstrates substantial deviation, whereas the proposed controller consistently maintains accurate trajectory tracking.

The proposed decoupled attitude controller demonstrated significant improvement in aligning the thrust axis with the guidance-required acceleration vector, especially in scenarios where large roll rotations ( $\phi_{cmd}$ ) were required to meet sensor orientation constraints. Unlike the classical controller, which sometimes induced unintended lateral motion due to global shortest-path which quaternion takes, the proposed method preserved guidance alignment while independently regulating roll attitude. 

\section{Conclusion}  \label{sec:conc}
This paper presented a novel quaternion-based decoupled thrust-axis attitude control strategy for  Chandrayaan-3. The approach addresses the inherent coupling present in classical quaternion feedback laws, particularly under large rotation demands about the thrust axis. By analytically separating thrust-axis and lateral-axis control, the proposed controller ensures guidance demand is met with shortest path ensuring minimal guidance and control interactions while simultaneously satisfying mission-specific sensor orientation requirements.

Simulation results using a high-fidelity autonomous trajectory simulator demonstrated that the decoupled controller effectively mitigates guidance-control interaction, leading to improved stability and trajectory accuracy during powered descent. Monte Carlo analyses under varying mission constraints $\phi_{cmd}$ further validated the robustness and flexibility of the proposed approach.

The proposed decoupled controller was implemented and rigorously tested through ground-based simulations, hardware-in-the-loop (HIL) setups, and field tests done using a Chandrayaan-3 equivalent ground test vehicle. These extensive validation efforts demonstrated the controller's effectiveness and robustness in meeting guidance objectives while accommodating thrust-axis rotation requirements. During the actual mission, the flight requirement for mission-specific thrust-axis rotation was set to zero, serving as a nominal baseline for performance assessment.

\section{Acknowledgement}  \label{sec:conc}
The authors would like to express their sincere gratitude to Mr. S. Sankaran, Director, Indian Space Research Organisation (ISRO), for providing the opportunity and motivation to pursue this work. Special thanks are extended to the Navigation, Guidance, and Control team for their continuous support and technical discussions throughout the development. The authors also gratefully acknowledge the expert review and constructive feedback provided by Dr. U.P. Rajeev, Dr. M.S. Siva, and Dr. Ravi Kumar, whose valuable insights greatly contributed to the technical rigor and clarity of this paper.

\bibliography{sample}
%\begin{thebibliography}{99}
%

%\bibitem{c1} Wieber, P.B., 2006, December. Trajectory free linear model predictive control for stable walking in the presence of strong perturbations. In 2006 6th IEEE-RAS International Conference on Humanoid Robots (pp. 137-142). IEEE.

%\end{thebibliography}

\end{document}